\documentclass[a4paper,fleqn,usenatbib,useAMS]{mnras}

\usepackage{graphicx}	
\usepackage{amsmath}	
\usepackage{amssymb}	
\usepackage{multicol}        
\usepackage{bm}		
\usepackage{pdflscape}	

\usepackage[T1]{fontenc}
\usepackage{ae,aecompl}
\usepackage{mathptmx}

\usepackage{hyperref}

\title[Environmental impact on disk galaxies]{The impact of environment on the lives of disk galaxies as revealed by SDSS-IV MaNGA}

\author[S. Zhou]{Shuang Zhou$^{1}$\thanks{Contact e-mail: \href{mailto:Shuang.Zhou@nottingham.ac.uk}{Shuang.Zhou@nottingham.ac.uk}},
Michael Merrifield$^{1}$,
Alfonso Arag{\'o}n-Salamanca$^{1}$,
Joel R. Brownstein$^{2}$,
\newauthor
Niv Drory$^{3}$,
Renbin Yan$^{4}$,
Richard R. Lane$^{5}$
\\
$^{1}$School of Physics \& Astronomy, University of Nottingham, University Park, Nottingham, NG7 2RD, UK\\
$^{2}$Department of Physics and Astronomy, University of Utah, 115 S. 1400 E., Salt Lake City, UT 84112, USA\\
$^{3}$McDonald Observatory, The University of Texas at Austin, 1 University Station, Austin, TX 78712, USA\\
$^{4}$Department of Physics, The Chinese University of Hong Kong, Shatin, N.T., Hong Kong, China\\
$^{5}$Centro de Investigaci{\'o}n en Astronomía, Universidad Bernardo O'Higgins, Avenida Viel 1497, Santiago, Chile\\}
\date{Last updated ???; in original form ???}

\pubyear{2022}

\begin{document}

\label{firstpage}
\pagerange{\pageref{firstpage}--\pageref{lastpage}}
\maketitle

\begin{abstract}
Environment has long been known to have significant impact on the evolution of galaxies, but here we seek to quantify the subtler differences that might be found in disk galaxies, depending on whether they are isolated, the most massive galaxy in a group (centrals), or a lesser member (satellites).  The MaNGA survey allows us to define a large mass-matched sample of 574 galaxies with high-quality integrated spectra in each category.  Initial examination of their spectral indices indicates significant differences, particularly in low-mass galaxies.  Semi-analytic spectral fitting of a full chemical evolution model to these spectra confirms these differences, with low-mass satellites having a shorter period of star formation and chemical enrichment typical of a closed box, while central galaxies have more extended histories, with evidence of on-going gas accretion over their lifetimes.  The derived parameters for gas infall timescale and wind strength suggest that low-mass satellite galaxies have their hot halos of gas effectively removed, while central galaxies retain a larger fraction of gas than isolated galaxies due to the deeper group potential well in which they sit.  S0 galaxies form a distinct subset within the sample, particularly at higher masses, but do not bias the inferred lower-mass environmental impact significantly. The consistent picture that emerges underlines the wealth of archaeological information that can be extracted from high-quality spectral data using techniques like semi-analytic spectral fitting.   
\end{abstract}



\section{Introduction}
Galaxies are complex systems, and their formation and evolution are affected by a variety of processes. In the current paradigm, galaxies form at the centre of dark matter halos whose formation and evolution are described by the Lambda cold dark matter cosmology \citep{White1978}. Baryonic matter accretes into these dark matter halos, fueling star formation. During this initial gas accretion and subsequent evolution, galaxies can experience a wide range of interactions with their environment, on scales ranging from mergers with individual galaxies to hydrodynamic interactions with galaxy clusters and the cosmic web. 

It is therefore perhaps unsurprising that galaxy properties depend on the environment that they inhabit. Most famously, high-density regions such as groups and clusters contain a significantly higher fraction of red galaxies, often associated with earlier morphological types \citep[e.g.][]{Oemler1974,Dressler1980,Postman1984}. With the advent of large scale surveys such as the Sloan Digital Sky Survey (SDSS,\citealt{York2000}), the environmental dependence of the galaxies' star formation histories has 
been extensively explored. Such surveys have revealed that satellite galaxies in groups and clusters  \citep[e.g.][]{Pasquali2010,Peng2012,Wetzel2012,Wetzel2013} and galaxies that live in high-density regions \citep[e.g.][]{Kauffmann2003,Baldry2006} are more likely to have their star formation quenched than isolated galaxies. Other properties related to the galaxies' star-formation histories show related environmental dependence: galaxies are redder \citep{Blanton2005ApJ,Li2006} and have a higher D4000 index \citep{Kauffmann2004,Li2006} in denser environments. Further observations have shown that this environmental dependence may persist to high redshifts \citep[e.g.][]{Peng2010,Darvish2016,Kawinwanichakij2017}. 
More recently, spatially-resolved observations allow the investigation of such environmental effects within the galaxies themselves, although several studies indicate that the environment seems to work mainly globally, with very little effect on the radial profiles of stellar population properties \citep[e.g.][]{Zheng2017,Spindler2018}.

There is some evidence that, in addition to affecting the star-formation activity of galaxies, the environment may also impact on their chemical evolution, although the effects seem more subtle, complex, and uncertain.  Early-type galaxies in low-density environments are more metal rich than their counterparts in high-density environments \citep{Sanchez2006}. Further, \cite{Pasquali2010} found that satellite galaxies seem to be more enriched in metals than central ones, especially in low-mass systems. However, the opposite trend was found by \cite{Zheng2017}: in large-scale sheets and voids, satellite galaxies are relatively more metal poor than central ones. \cite{Zheng_etal2019} also found that the environment appears to have a modest effect on the $\alpha$-element abundances of galaxies, especially at low and average galaxy masses. Finally, as with stellar ages, although the global metallicity of galaxies correlates with their environment, metallicity gradients show weak or no environmental dependence \citep[e.g.][]{ Zheng2017,Goddard2017}.

In trying to understand the physical drivers of these differences, it is apparent that varying processes are at work depending on whether a galaxy is central within a halo or a satellite system. Indeed, analysing SDSS data, \cite{Peng2012} concluded that the quenching of satellite galaxies in galaxy groups provides the starting point of environmental effects. \cite{Kovac2014} further confirmed the dominant role of satellite quenching over time by analysing samples from zCOSMOS-bright data; they found little evidence that the red fraction of central galaxies depends on over-density, while satellites are commonly redder at all over-densities, at least back to $z\sim0.7$.

Currently, it is still unclear what distinction in physical mechanism is responsible for these differences in what is termed "environmental quenching". A range of processes have been suggested that can contribute, which all involve depriving a galaxy of the cold gas that it needs to keep forming stars. For example, a galaxy may experience ram-pressure stripping when it moves through a hot gaseous medium, which can rapidly remove the fuel for star formation and quench the galaxy \citep[e.g.][]{Gunn1972,Abadi1999}. Alternatively, if only the gas in the outer part of the galaxy is removed, or is prevented from falling to the galaxy through heating, the galaxy would experience `strangulation' and gradually quench when the existing fuel is exhausted \citep[e.g.][]{Larson1980,Balogh2000}. Tidal interactions between galaxies in a group or between galaxies and the dark matter halo of the group or cluster can also cause stripping of the gas \citep[e.g.][]{Read2006}. In each case, the net effect is the same, so the observable consequences of quenching will be similar, making it difficult to determine which processes are important in any particular situation.  However, given the physical differences in these processes, we might expect to be able to distinguish between them on the basis of subtler indicators such as the timescale over which star formation shuts down, and the way that gas loss can alter a galaxy's chemical evolution.  

Fortunately, we are now in a position to investigate these subtler factors to distinguish between such effects.  The recently-completed Mapping Nearby Galaxies at Apache Point Observatory (MaNGA) survey \citep{Bundy2015} has provided us with high-quality spectra from a well-defined sample of over 10,000 nearby galaxies, which allows us to select a large set of galaxies in different environments, to search for quite subtle variations. The size of the survey also means that we can focus on a homogeneous subset of galaxies, matched in mass between environments, and chosen to have a shared underlying morphology -- in this case, all disk galaxies -- to seek to minimize the variations that might be expected for galaxies of very different types in different environments. Moreover, we have recently shown that, by fitting relatively simple stellar and chemical evolution models directly to the high-quality spectra that MaNGA provides, it is possible to determine physical parameters related to the timescales of star formation and rates of gas accretion and loss that should allow us to distinguish between quenching mechanisms (\citealt{zhou2022}, hereafter Paper~I).  

To this end, we have set out the analysis in this paper as follows.  The data and sample selection, and a preliminary look at the evidence it presents for environmental influence, are described in \S\ref{sec:data}.  The "semi-analytic spectral fitting" process, by which we fit a simple physically-motivated evolutionary model directly to the spectra, is summarized in \S\ref{sec:analysis}. The inferred evolutionary parameters for central, satellite and isolated galaxies, and their implications for the drivers of environmental quenching, are described in \S\ref{sec:results}. Finally, we summarise the key findings in \S\ref{sec:summary}.


\section{Data}
\label{sec:data}
\subsection{The MaNGA Survey}
As part of the fourth generation of SDSS (SDSS-IV, \citealt{Blanton2017}), MaNGA has completed its mission to collect spatially-resolved high-quality spectra for a well-defined sample of over 10,000 nearby galaxies (redshift $0.01<z<0.15$, \citealt{Yana2016,Wake2017}). MaNGA targets were selected to span the stellar mass range $5\times10^8 h^{-2}{M}_{\odot} \leq M_*\leq 3 \times 10^{11} 
h^{-2}{M}_{\odot}$ \citep{Wake2017}, using  stellar masses from the NASA Sloan Atlas catalogue \footnote{\label{foot:nsa}\url{ http://www.nsatlas.org/}} (NSA, \citealt{Blanton2005}). Using two dual-channel BOSS spectrographs \citep{smee2013} mounted on the 2.5\,m telescope \citep{Gunn2006}, intermediate resolution ($R\sim2000$, \citealt{Drory2015}) spectra that cover $3600-10300${\AA} in wavelength were obtained out to at least 1.5 effective radii for all the target galaxies \citep{Law2015}. The raw spectra were reduced and calibrated \citep{Yanb2016} via the Data Reduction Pipeline (DRP; \citealt{Law2016}) to produce science-ready spectra with flux calibrations batter than $5\%$ across most of the wavelength range. The Data Analysis Pipeline (DAP;\citealt{Westfall2019,Belfiore2019}) processed these reduced data and provided the measurements of stellar kinematics, emission-line properties and spectral indices employed in this work.

\subsection{Sample selection and data reduction}
\begin{figure}
    \centering
    \includegraphics[width=0.45\textwidth]{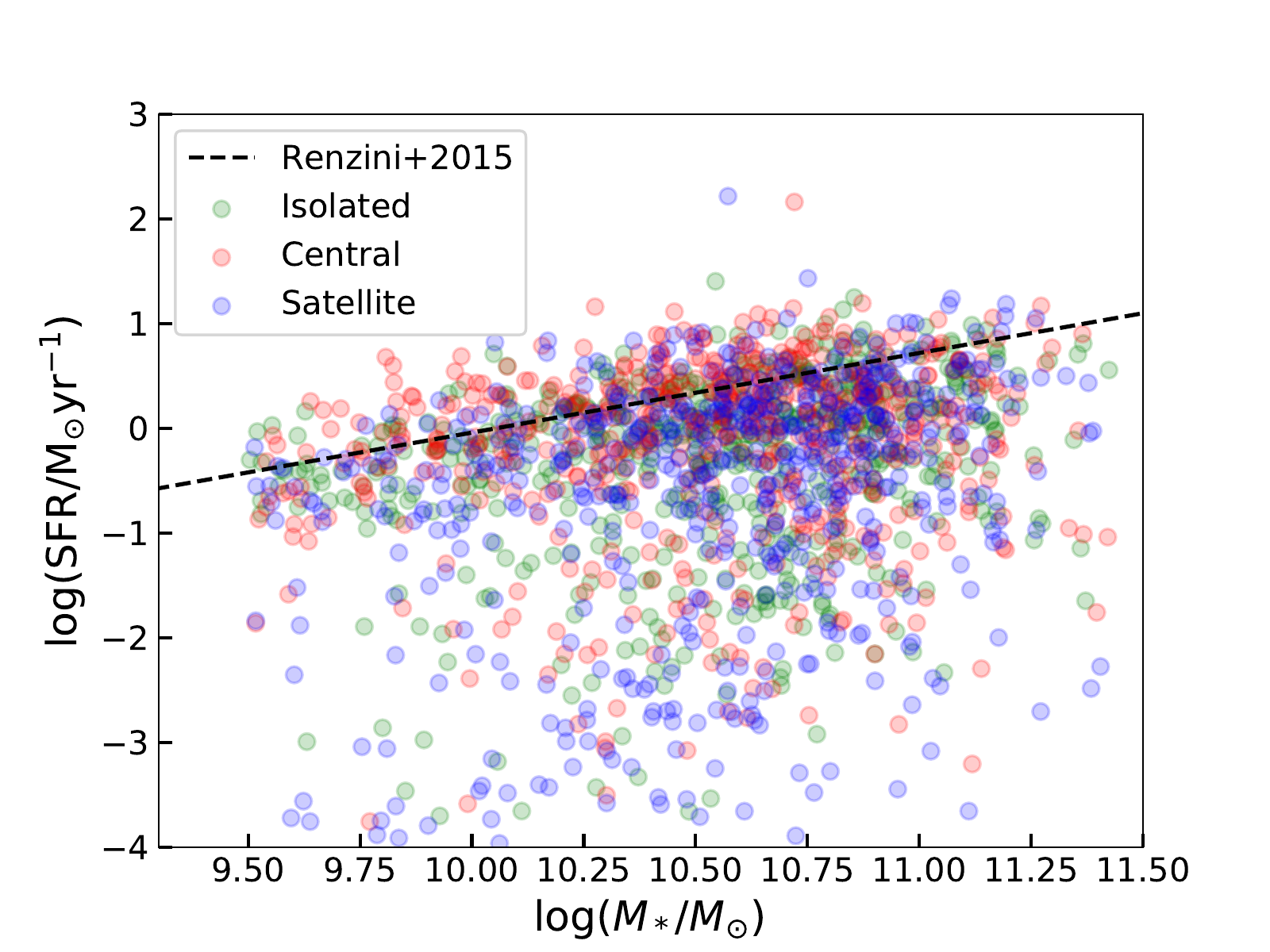}
     \caption{The star formation rate  as a function of stellar mass for our sample galaxies. Isolated, central and satellite galaxies are shown in green, red and blue, respectively. The black dash line shows the star forming main sequence obtained by \citet{Renzini2015} as reference.}
     \label{fig:SFR}
\end{figure}

The intent of this paper is to investigate how the stellar population properties in galaxies depend on the environment that they inhabit, specifically whether they are central or satellite galaxies in groups, or are isolated objects. To identify such systems, we make use of the 11th MaNGA Product Launch (MPL11) data, which comprises the full 10,010 unique galaxies observed, and was released with the SDSS-IV Data Release 17 (DR17, \citealt{SDSSDR17}). We are also looking to select a sample in which mergers are unlikely to have confused the evolutionary picture significantly, so focus on disk-like systems \citep{Mo1998}. To this end,
we make use of the value-added catalogue (VAC), the MaNGA Morphology Deep Learning DR17 Catalogue \citep{Sanchez2022}, which provides classifications for all the MaNGA MPL11 galaxies. In this catalogue, several indicators are provided to describe the morphology of the galaxy. $\rm P_{LTG}$
is the probability of being late-type galaxies from the deep-learning approach. T-Type is another indicator with T-type$\leq$0 for early-type galaxies, T-type$>$0 for late-type galaxies, and T-type=0 for S0s. Visual class (VC) comes from visual inspection of the image of the galaxy, in which late-type galaxies are marked with VC=3 and S0 with VC=2. In addition, a visual classification flag (VF) is used to indicate the robustness of the visual classification, in which certain visual classification results have VF=0. There are multiple ways in which these indicators can be combined to create a single measure of morphology; here, we follow the criteria suggested in \cite{Sanchez2022} to select disk galaxies as those in the VAC labeled with 
\begin{itemize}
    \item spiral: ($\rm P_{LTG}>$0.5) and (T-Type$>$0) and (VC=3) and (VF=0),
    \item S0: ($\rm P_{LTG}<$0.5) and (T-Type$<$0) and (PS0$>$0.5) and (VC=2) and
(VF=0),
\end{itemize}
which yields 5125 spirals and 891 S0s. To ensure the robustness of morphological classifications and to minimize dust attenuation effects, we also require the sample galaxies to be reasonably face-on; after applying an axis-ratio cut of $b/a>0.5$, we obtain a sample of 4416 reasonably face-on disk galaxies.

To obtain the requisite environment information, we make use of another VAC, the Galaxy Environment for MaNGA Value Added Catalog (GEMA-VAC). In this VAC, the environment information of MaNGA galaxies is derived using the methods described in \cite{Argudo2015}, \cite{Etherington2015} and \cite{Wang2016}. The final catalogue will be described in more detail in Argudo-Fern\'adez et al. (in prep.), but the VAC itself has been released with SDSS DR17. 
We use the group environment of MaNGA galaxies provided in the VAC, in which galaxies are allocated to groups using the catalogue of \cite{Yang2007}. Using this information, we divide the sample into three classes:  
\begin{itemize}
    \item isolated: number of galaxies NG=1 in the group
    \item central: the most massive galaxy in a group with NG$>1$
    \item satellite: not the most massive galaxy in a group with NG$>1$.
\end{itemize}

One further issue is that galaxies selected in different environments could have systematically different properties.  In particular, their mass distributions are likely different, with more massive systems found in higher-density regions.  Since mass is likely the biggest single intrinsic driver of evolutionary differences, this effect would mask the extrinsic environmental effects that we are looking for. To mitigate this issue, we have mass-matched the sub-samples of galaxies in each environmental category in the stellar mass range $9.5<\log(M_*/{\rm M}_{\odot})<11.5$. This range is divided into bins of width $\log(M_*/{\rm M}_{\odot})=0.1$, which is approximately equal to the uncertainty in the mass measurement, and within each bin we select all $N$ galaxies from the least-popular category, and a random sub-sample of $N$ galaxies from the other two categories.  In this way, we construct a mass-matched sample in the three categories, which ends up with 574 galaxies in each environmental class.

In this sample, most of the central galaxies come from relatively small groups, with a median number of group members NG=3, and 90\% of the groups have NG<10. Satellite galaxies come from bigger groups with a median NG=16. In addition, the GEMA-VAC provides halo mass estimations for MaNGA galaxies obtained from \cite{Yang2007}. We use these halo masses to characterise in more detail the dark matter halo properties of the sample galaxies. Isolated galaxies have halo masses in the range
$10^{11.5}<M_{\rm h}/{\rm M}_{\odot}<10^{13}$ with a median of $10^{12.4}{\rm M}_{\odot}$. Central galaxies generally reside in relatively larger dark matter halos with halo masses in the range $10^{11.5}<M_{\rm h}/{\rm M}_{\odot}<10^{14}$ and a median of $10^{12.6}{\rm M}_{\odot}$. As might be expected for satellite galaxies that do not define the mass of the halo in which they reside, the range for these systems is much larger, spanning from $10^{12}{\rm M}_{\odot}$ to $10^{14.5}{\rm M}_{\odot}$ with a median of $10^{13.7}{\rm M}_{\odot}$. In what follows we will explore how this range of environments has affected the evolution of the galaxies they host.

\subsection{Environmental dependence of the galaxies' empirical properties}
\label{subsec:empirical}

Before getting into the detailed modelling that will tell us about the physical processes involved in environmental quenching, it is helpful to look for clues in the simpler empirical data available directly from the MaNGA DAP.  In \autoref{fig:SFR} we present the star formation activity of the sample galaxies as a function of their masses. For this analysis, we use H$\alpha$ flux measurements provided by MaNGA DAP (where detected) to estimate the star formation rate. Using the Calzetti extinction law \citep{Calzetti2000} and assuming a intrinsic H${\alpha}$/H${\beta}$ ratio of 2.87 \citep{Osterbrock2006}, we correct the H$\alpha$ flux for dust attenuation. A conversion factor from \cite{Murphy2011} calibrated with a Chabrier \citep{Chabrier2003} IMF,
\begin{equation}
\rm \frac{SFR}{  M_{\odot}yr^{-1}}=5.37\times10^{-42}\frac{L(H\alpha)}{erg\  s^{-1}},   
\end{equation}
is then applied to calculate the SFR, which is shown as a function of the stellar mass from the NSA. From the plot, we see that most of the sample galaxies lie close to the star-forming main sequence. However, significant deviations from the main sequence towards lower SFRs are also seen in many galaxies, particularly the satellites systems. Clearly, this offers evidence that the present-day star-formation rates in disk galaxies are significantly affected by environment.

\begin{figure}
    \centering
    \includegraphics[width=0.45\textwidth]{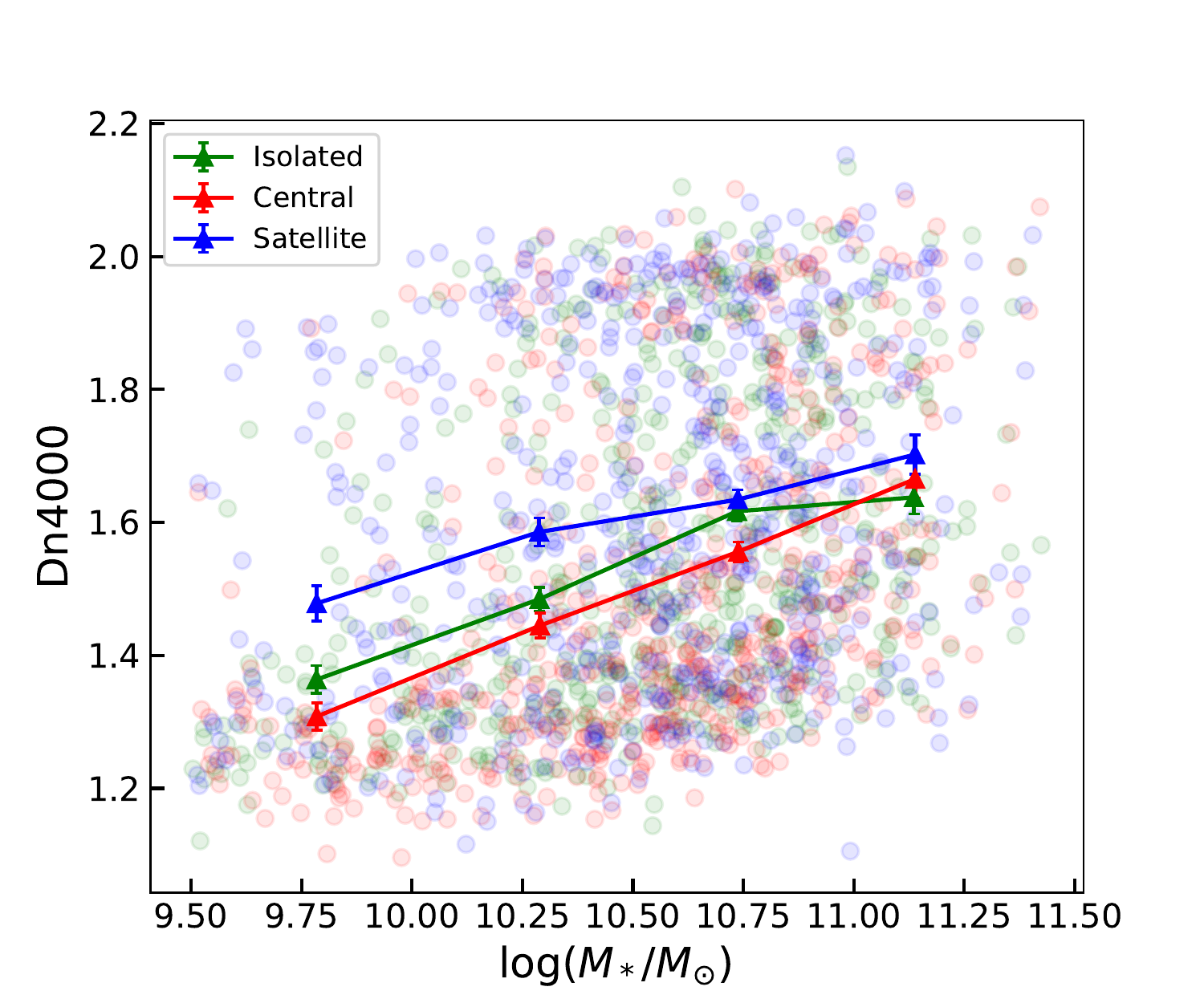}\\
    \includegraphics[width=0.45\textwidth]{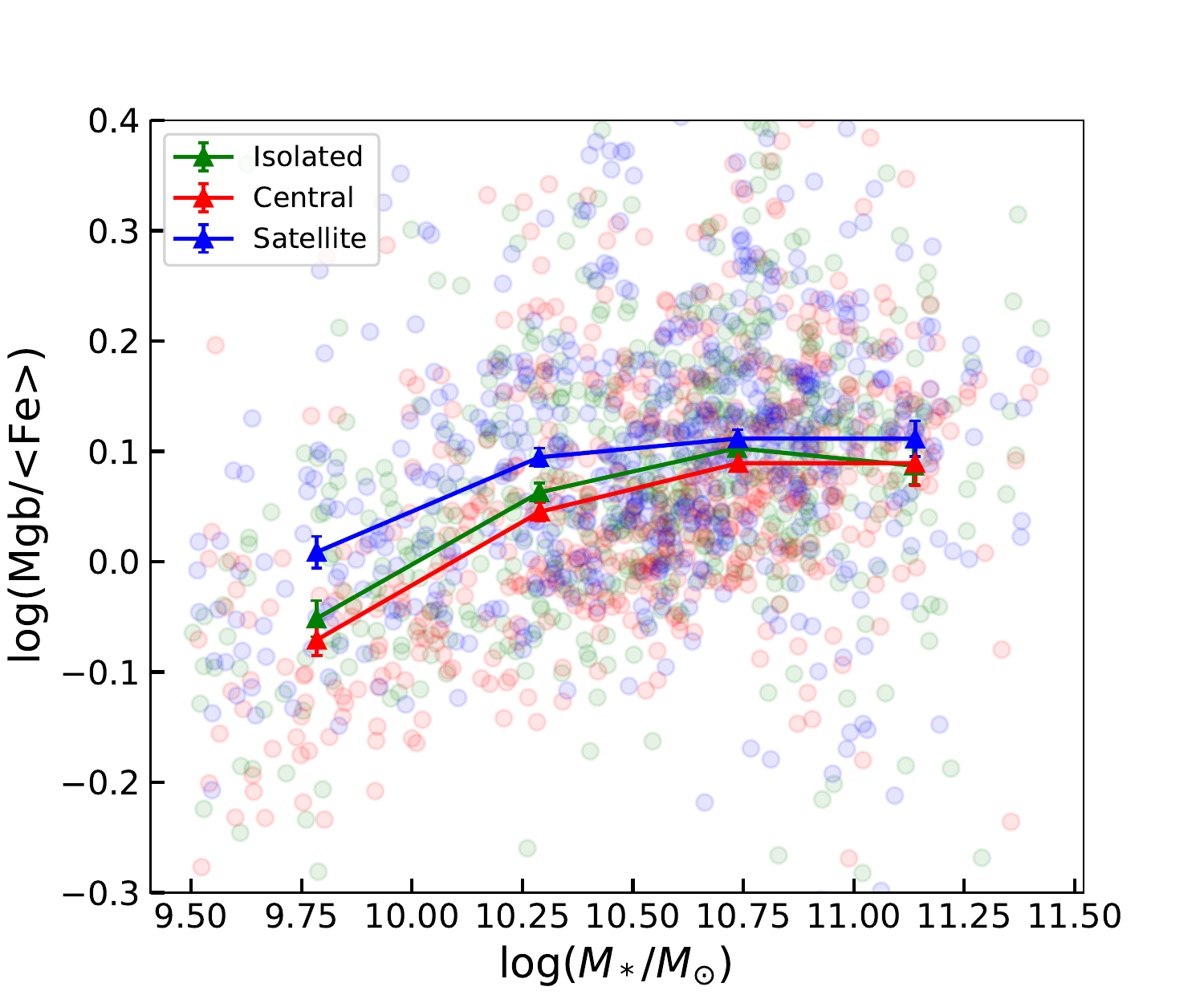}    
     \caption{Dn4000 (top) and $\rm Mgb/\langle Fe\rangle$ (bottom) as functions of stellar mass for our sample galaxies. Isolated, central and satellite galaxies are shown in green, red and blue, respectively. Triangles linked by lines are mean results in different stellar mass bins, with error bar showing the error estimated from the jackknife resampling method. }
     \label{fig:index}
\end{figure}

To probe further back into the star-formation histories of galaxies of different types, we can look for empirical evidence in various absorption-line indices that have been measured for these galaxies.  In particular, the Dn4000 index is often used as an indicator of the stellar population, and specifically a value of Dn4000$<1.6$ will be an indicator of star formation within the past 1--2\,Gyr \citep[e.g.][]{Kauffmann2003}.  In addition, the $\rm Mgb/\langle Fe\rangle \equiv Mgb/(0.5*Fe5270+0.5*Fe5335)$ index is a proxy for the relative abundance of alpha elements, for which higher values are found in galaxies whose stars formed on a shorter timescale \citep[e.g.][]{Worthey1994,Thomas2005,Zheng_etal2019}.  \autoref{fig:index} shows the average values for these indices, integrated out to the effective radius of all the sample galaxies, as a function of their stellar masses.  Although there is a good deal of scatter, the average lines on these plots show clear systematic differences between galaxies in different environments.  The satellite galaxies show stronger values of both indices, indicating relatively little recent star formation, consistent with what we have learned from H$\alpha$, and a shorter timescale for their formation.  Interestingly, though, we also see consistent, if subtle, differences between central and isolated galaxies, even in these relatively crude measures of galaxy evolution, which motivates a more through investigation of its causes.

\section{Analysis}
\label{sec:analysis}
\subsection{The chemical evolution model}
\label{subsec:model}

In order to investigate the physical processes that underlie the spectral differences between galaxies in different environment, we fit a chemical evolution model directly to the spectra.  The model is designed to be sufficiently general to capture the main physical processes in a wide variety of formation scenarios, but simple enough to produce robust fits to the spectra.  This semi-analytic spectral fitting approach is described in detail in Paper~I, but we summarize the main steps in the approach here.  

To construct a sufficiently-general picture of the star-formation and chemical evolution of a galaxy, we have to take into account gas inflow and outflow processes as well as the star formation triggered by the resulting gas reservoir. We therefore consider a model in which the gas mass evolution of the galaxy can be described by
\begin{equation}
\dot{M}_{\rm g}(t)=\dot{M}_{\rm in}(t)-\psi(t)+\dot{M}_{\rm re}(t)-\dot{M}_{\rm out}(t).
\end{equation}
The first term characterises the gas inflow, which is assumed to have an exponentially decaying form 
\begin{equation}
\dot{M}_{\rm in}(t)=A e^{-(t-t_0)/\tau}, \ \ \ t > t_0.
\end{equation}
The second term represents the gas lost to star formation, with the third term accounting for the mass ejection from dying stars, and the final term describing mass lost from the system. We assume a linear Schmidt law \citep{Schmidt1959} to model the star formation activity, so
\begin{equation}
\psi(t)=S\times M_{\rm g}(t),
\end{equation}
where $S$ is the star-formation efficiency. We assume this efficiency to be constant throughout the evolution and estimate it with the extended Schmidt law
proposed by \cite{Shi2011}, 
\begin{equation}
\label{eq:sfe}
    S (yr^{-1})=10^{-10.28\pm0.08} \left( \frac{\Sigma_*}{M_{\odot} pc^{-2}} \right)^{0.48},
\end{equation}
where $\Sigma_{*}$ is the stellar mass surface density of the galaxy. We obtain an approximate average value for  $\Sigma_{*}$ from the current stellar mass and effective radius in the NSA, $\Sigma_*=0.5\times M_*/(\pi R_{\rm e}^2)$. 

In this model, a constant mass return fraction of $R=0.3$ is assumed, so that the stars formed in each generation will return 30\% of their stellar masses to the ISM. For the mass-loss term, we follow the canonical assumption that the outflow strength is proportional to the star formation activity:
\begin{equation}
\label{eq:outflow}
\dot{M}_{\rm out}(t)=\lambda\psi(t),
\end{equation}
where the dimensionless quantity $\lambda$ is the relative outflow strength, often called the `wind parameter.'

In Paper~I, we assumed the outflow to be constant, but with a sharp cut-off at some time (to represent when a galaxy's potential well became too deep for the winds to escape).  This assumption allows a secondary star-formation episode during the evolution of the galaxy, but the sudden cutoff leads to discontinuity in the chemical and star-formation evolution.  Here, we consider a somewhat gentler model for the changes in the wind parameter, which is assumed to vary linearly between an initial value $\lambda_{\rm b}$ at the beginning of galaxy formation (14Gyr ago,) and a  value of $\lambda_{\rm e}$ at the present day:
 \begin{equation}
 \label{eq:ouflow}
\lambda(t)=\lambda_{\rm b}+(\lambda_{\rm e}-\lambda_{\rm b})t/(14\,{\rm Gyr}).
\end{equation}
We found that this simple refinement avoids the likely-unphysical discontinuity in properties, while leaving the statistical results largely unchanged. Using all those ingredients, the equation that characterise the gas mass evolution can be written as
 \begin{equation}
\label{eq:massevo}
\dot{M}_{\rm g}(t)=
   A e^{-(t-t_0)/\tau}-S(1-R+\lambda(t)) M_{\rm g}(t).
\end{equation}

For the chemical evolution, we adopt the usual instantaneous mixing approximation, in which the gas in a galaxy is always well mixed during its evolution. The equation that characterize the chemical evolution can
then be written as
\begin{equation}
\label{eq:cheevo}
\begin{aligned}
\dot{M}_{Z}(t)=& Z_{\rm in}Ae^{-(t-t_0)/\tau}-Z_{\rm g}(t)(1-R)SM_{\rm g}(t) +\\
& y_Z(1-R)SM_{\rm g}(t) -Z_{\rm g}(t)\lambda(t) SM_{\rm g}(t),
\end{aligned}
\end{equation}
where $Z_{\rm g}(t)$ is the gas phase metallicity and $M_{Z}(t)\equiv M_{\rm g}\times Z_{\rm g}$. The first term is the inflow, which brings in gas with metallicity $Z_{\rm in}$, which is simply assumed to be $Z_{\rm in}=0$ throughout this work. The second term characterizes the metal locked up in long-lived stars, while the third term represents the metal-enriched gas returned by dying stars. The last term characterise how the outflow blows away metal-enriched gas, with a time-dependent outflow strength specified by \autoref{eq:ouflow}.

\subsection{Fitting the model to observed data}
\label{subsec:spectral_fitting}
Now that we have a very general model for the star-formation history (SFH) and chemical evolution history (ChEH) of a galaxy, we are in a position to constrain its parameters using the spectral data from each galaxy.  We do so by adopting a Bayesian framework, using an updated version of the Bayesian Inference of Galaxy Spectra ({\tt BIGS}), as presented in Paper I. In short, a set of model parameters are generated from an appropriate prior distribution (listed in \autoref{tab:paras}), which are used to calculate the SFH and ChEH following \autoref{eq:massevo} and \autoref{eq:cheevo}. The present-day gas phase metallicity $Z_{\rm g}$ is obtained from the ChEH. Using the \cite{BC03} stellar population models, we calculate a model spectrum corresponding to the SFH and ChEH following the standard stellar population synthesis approach (see review by \citealt{Conroy2013}). We use \cite{BC03} models constructed with the `Padova1994' isochrones, the Chabrier \citep{Chabrier2003} IMF and the STILIB \citep{Borgne2003} stellar templates, which covers metallicities from $Z =0.0001$ to $Z = 0.05$, and ages from $0.0001\,{\rm Gyr}$ to $20\,{\rm Gyr}$. The library models have a FWHM resolution of 3\AA\ in the wavelength range $3200$ -- $9500$\AA, which makes them well-suited to fitting MaNGA spectra observed at $3600$ -- $10300${\AA} with low redshifts ($z\sim0.03$, \citealt{Wake2017}), but we must also broaden the stellar templates to match the broadening due to stellar velocity dispersion and instrumental effects, which we infer from an initial {\tt pPXF} fit to the spectra \citep{Cappellari2017}.  We can then compare the inferred model spectrum to the MaNGA data and vary the parameters to find the best-fit physically-motivated model.

This method was extensively tested in Paper~1, in which we showed that, with spectral data at a sufficient signal-to-noise ratio (SNR), it can reliably reproduce the overall SFH and ChEH of galaxies even when the simplified model does not exactly match the detailed history of a galaxy.  However, it does require quite a high SNR to fit reliably. Fortunately, here we are interested in the global properties of galaxies, so, as described in Paper~1, can combine all the data within 1$R_{\rm e}$ of each galaxy to obtain a spectrum with a SNR of $\sim 70$ per {\AA}, which is sufficient to robustly reproduce the system's history.

In Paper~I, we used emission-line diagnostics to further constrain the present-day star-formation rate and gas-phase metallicity.  In this work, however, as we no longer focus only on spiral galaxies, some objects do not offer these constraints.  However, we can use a population-wide relation instead, at least for the metallicity constraint.  Specifically, \cite{Andrews2013} demonstrated that there is quite a tight relationship between mass and gas-phase metallicity, so we can assign an expected gas-phase oxygen abundance using
\begin{equation}
12+{\rm log(O/H)}=8.798-{\rm log}\left(1+
\left(\frac{10^{8.901}}{M_\star/{\rm M}_{\odot}}\right)^{0.64}\right).
\label{eqn:metal}
\end{equation}
This oxygen abundance is then converted to total metallicity using a solar metallicity of 0.02 and oxygen abundance of $12+\log({\rm O/H})=8.83$ \citep{Anders1989}. We note that this relation is chosen to be consistent with the calibration of the \citet{BC03} SSP models; alternative calibrations can also be applied but appropriate caution is urged when comparing with our results. The likelihood function used in deriving a best fit can then be written as:

\begin{equation}
\label{likelyhood}
\ln {L(\theta)}\propto-\sum_{i}^N\frac{\left(f_{\theta,i}-f_{\rm D,i}\right)^2}{2f_{\rm err,i}^2}-
\frac{(Z_{\rm g,\theta}-Z_{\rm g,D})^2}{2\sigma_{Z}^2
},
\end{equation}
where $f_{\theta, i}$ is the flux predicted from the model with parameter set $\theta$, while $f_{\rm D, i}$ is the observed flux from the stacked spectrum. $f_{\rm err,i}$ represents the corresponding error spectrum, and the sum is made over all $N$ wavelength points. $Z_{\rm g,\theta}$ is the current gas phase metallicity from the model, while $Z_{\rm g,D}$ is the gas phase metallicity estimated from the empirical relation of \autoref{eqn:metal}. $\sigma_{Z}$ is the uncertainty in the gas phase metallicity estimates, which was set to $\sigma_{Z}=0.1Z_{g,D}$ \citep{zhou2022}.

\begin{table}
	\centering
	\caption{Priors of model parameters used to fit galaxy spectra}
	\label{tab:paras}
	\begin{tabular}{lccr}
		\hline
		Parameter & Description & Prior range\\
		\hline
		$y_Z$ & Yield parameter& $[0.0, 0.08]$\\
		$\tau$ & Gas infall timescale & $[0.0, 14.0]$Gyr\\
		$t_{0}$ & Start time of gas infall & $[0.0, 14.0]$Gyr\\
		$\lambda_b$ & The wind parameter $14\,$Gyr ago & $[0.0, 10.0]$\\
		$\lambda_e$ & The current wind parameter  & $[0.0, 10.0]$\\
		$E(B-V)$&  Dust attenuation parameter & $[0.0, 0.5]$\\
		\hline
	\end{tabular}
\end{table}

\begin{figure*}
    \centering
    \includegraphics[width=0.95\textwidth]{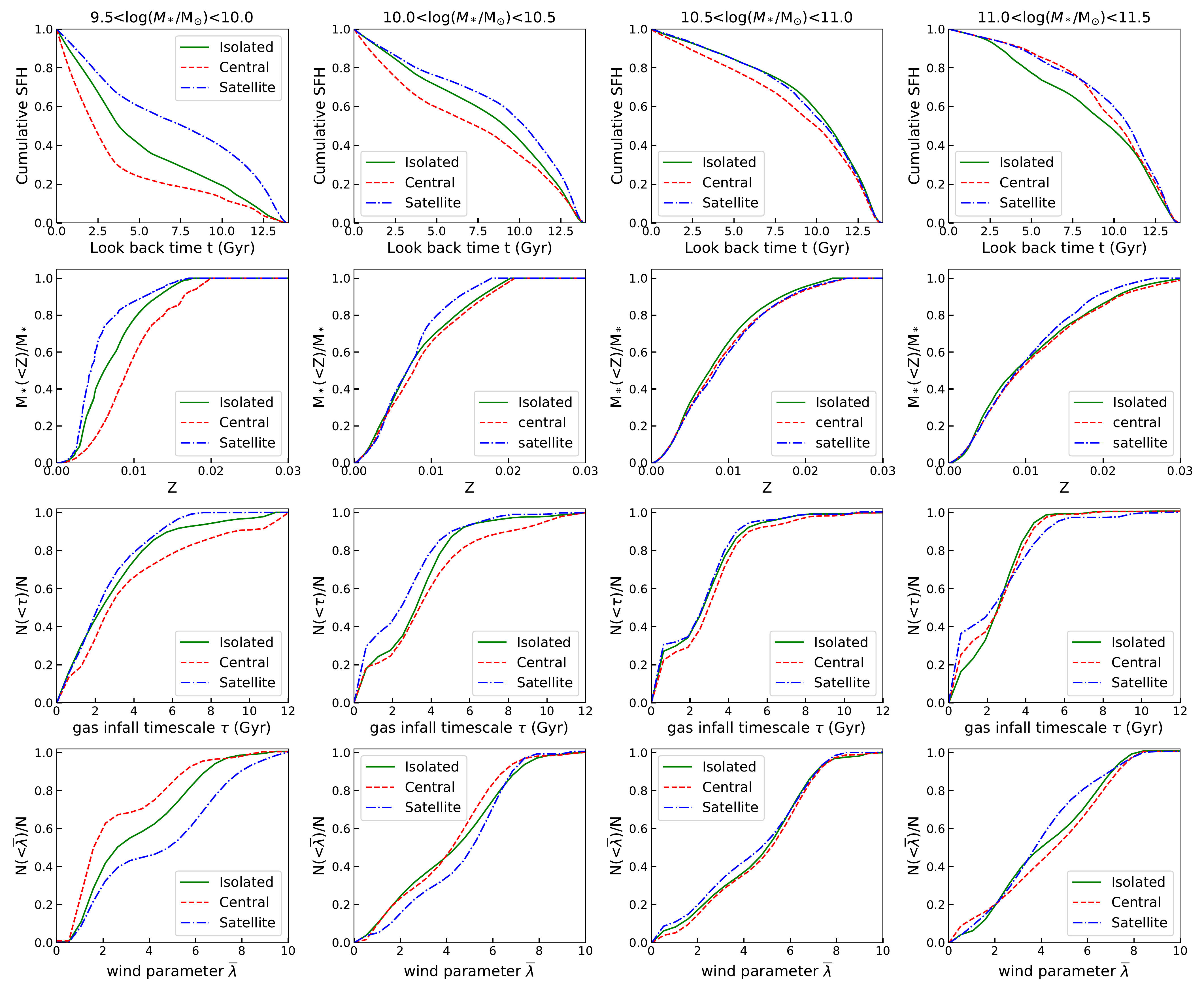}
 
     \caption{Environment dependence of galaxy properties derived from best-fit models for our sample galaxies.
     The first row shows the cumulative SFH while the second row shows the cumulative metallicity distribution, averaged over galaxies in each bin and category respectively. To interpret the derived mass and chemical evolution, the third and fourth panel show the cumulative distributions of the gas infall time-scale $\tau$ and mass-weighted averaged wind parameter $\overline{\lambda}$, as indicators of gas infall and outflow properties respectively in our sample galaxies. In each row, from left to right, different panels show results from different stellar mass bins as indicated. In each panel, green, red and blue lines are results obtained from isolated, central and satellite galaxies respectively. }
     \label{fig:gal_envir}
\end{figure*}

\section{Results and discussion}
\label{sec:results}
Having performed the semi-analytic spectral fitting on the full mass-matched sample of central, satellite and isolated galaxies, we are now in a position to quantify and interpret, at least in broad terms, the empirical differences in their properties found in \autoref{subsec:empirical} in terms of the basic physical parameters that this analysis provides.  In \autoref{fig:gal_envir}, we present the average derived properties and underlying physical parameters for the mass-matched samples of isolated, central and satellite galaxies, divided into four mass bins.  We present these properties in cumulative form, in part because in some cases it has a simpler physical interpretation \citep[e.g.][]{Greener2021}, but also because it makes clearer the systematic differences between the sub-samples.

In terms of star-formation history (top row of \autoref{fig:gal_envir}), it is clear that in the lowest stellar mass bin ($10^{9.5}<M_*/{\rm M}_{\odot}<10^{10.0}$), central galaxies accumulate their stellar mass significantly later than satellite galaxies: on average less than 30\% of the stellar mass in central galaxies formed earlier than 5\,Gyrs ago, while in the same period satellite galaxies had accumulated more than 60\% of their stellar mass. The evolutionary history of isolated galaxies lie between these two extremes. These SFHs are in line with indications from the lower panel of \autoref{fig:index}, in which the higher $\alpha$ element abundance in low-mass satellites flags their shorter star formation timescales. The difference between central and satellite galaxies becomes weaker in higher mass galaxies, and in galaxies with  $M_*/{\rm M}_{\odot}>10^{10.5}$ the difference is almost negligible. The different formation history of central and satellite galaxies, as well as its variation with stellar mass, is consistent with previous investigations \citep[e.g.][]{Pasquali2010,Wetzel2012,Peng2012}, but the evolutionary model fitted here means that we will be able to go on to attribute a physical interpretation to these differences.

The second row of \autoref{fig:gal_envir} shows the associated average cumulative metallicity distribution (CMDF) in the different mass bins.  Once again, it is the low mass galaxies that display the biggest differences, with satellite galaxies having $\sim80\%$ of stars with $Z<0.005$, while in central galaxies of similar stellar mass this fraction decreases to $\sim20\%$. The low-metal-rich CMDF of low-mass satellite galaxies is characteristic of the so-call `closed-box' or `leaky-box' model, in which there is no significant gas infall to the system \citep{Tinsley1974}; in contrast, the CMDFs in central galaxies are closer to the predictions of `accreting box' models \citep{Tinsley1974,Tinsley1980}, in which a replenishing supply of new gas is provided, which suppresses  the relative fraction of low-metallicity stars. This effect is still present in higher mass bins, but is much reduced, indicating that environment is a much less important factor in the chemical evolution of these galaxies.

We can now quantify plausible drivers of these differences using the physical parameters inferred in the semi-analytic spectral fitting process.   The third row of \autoref{fig:gal_envir} shows one of the primary drivers of differences between galaxies, the gas infall timescale, as a function of both environment and mass. It is apparent that, at low and intermediate masses, central galaxies tend to have longer gas infall timescales, while the gas infall occurs over a systematically shorter period in satellite galaxies. In the lowest mass bin ($10^{9.5}<M_*/{\rm M}_{\odot}<10^{10.0}$), almost all of the satellite galaxies have gas infall timescale less than  $\sim$5\,Gyr, while around $40\%$ of central galaxies have timescales greater than $\sim$5\,Gyr, with isolated galaxies somewhere in-between. The median gas infall timescales for satellite, isolated and central galaxies are 2.6, 2.9 and 3.9\,Gyr, respectively. In more massive galaxies ($M_*/M_{\odot}>10^{10.5}$) these differences become less systematic: the median gas infall timescales for all the three categories becomes shorter, and all converge to around 2.5\,Gyr in the highest mass bin ($10^{11}<M_*/{\rm M}_{\odot}$).
Such variations with stellar mass are in line with the down-sizing formation scenario, with timescales compressed for more massive objects. In addition, The environment effects found in low-mass galaxies fit with suggested physical mechanisms for quenching galaxy formation, in which satellites falling into a larger dark matter halo can experience the removal of their hot gaseous halo through tidal ``strangulation'' processes \citep{Balogh2000,Pasquali2010} or even the direct ram-pressure stripping of their cold star-forming gas \citep{Gunn1972,Abadi1999}, resulting in an early end to their gas supply.  The fitted timescales for gas depletion in satellites in \autoref{fig:gal_envir} are relatively long ($\sim3$\,Gyr), which argues for the gentler process of strangulation, consistent with the conclusions from entirely independent lines of evidence reached by \citet{vandenBosch2008}, \citet{Pasquali2010} and \citet{Peng2015}.

One other interesting phenomenon apparent in the distribution of $\tau$ in \autoref{fig:gal_envir} is the step in its cumulative distribution that is particularly apparent in more massive galaxies.  This feature represents a separate population of galaxies in which the gas infall timescale is less than $\sim 1$\.Gyr.  Further investigation of these galaxies reveals that they are almost all classified as S0 or lenticular galaxies.  These systems accumulated their stellar masses very early, so are barely resolved in the current analysis.  It is nonetheless interesting to note that they make up a higher fraction of satellite than central galaxies, again implying that the process responsible for their creation is dependent on environment.  

The other major component in the semi-analytic model is outflow, as described by the wind parameter $\lambda$.  Since this parameter is allowed to vary over the lifetime of each galaxy, we quantify it in each system using an average value that reflects when it will have most affected the star-formation history of the galaxy, 
\begin{equation}
 \overline{\lambda} \equiv \frac{\int_0^{t_{\rm u}} \psi(t) \lambda(t) dt}{\int_0^{t_{\rm u}} \psi(t) dt},    
\end{equation}
where $\psi(t)$ is the star formation rate at time $t$ and ${t_{\rm u}}=14.0$\,Gyr is the age of the Universe. The bottom row of \autoref{fig:gal_envir} shows how this average wind parameter varies with galaxy mass and environment. Once again, it is only in the lowest-mass galaxies where we see a significant effect, with satellite galaxies showing much stronger wind parameters than central ones, with isolated galaxies somewhere in between. The median average wind-parameter for satellite, isolated and central galaxies in the lowest mass bin ($10^{9.5}<M_*/{\rm M}_{\odot}<10^{10.0}$) are 4.5, 3.5 and 2.6 respectively, while in relatively massive galaxies ($10^{10}<M_*/{\rm M}_{\odot}$), the median values all converge to around 4. Although this parameter was formulated in terms of a star-formation-driven wind, it really just measures the effectiveness of the loss of processed gas from the galaxy.  The relatively high rate of loss from satellites therefore fits with a picture in which a system falling into a more massive halo might be expected to lose its own extended halo of hot processed gas through interaction with other group members \citep{Gunn1972,Abadi1999}. Similarly, the greater ability of a central galaxy to hold on to this processed gas when compared to an isolated system presumably reflects the deeper potential well in which it sits. The disappearance of difference between the wind-parameters in galaxies with stellar masses higher than ${10^{10.0}\rm M}_{\odot}$ implies that massive satellite galaxies do not lose more gas than centrals. This convergence in properties can likely be attributed to the greater ability of massive galaxies to hold on to processed gas against external influences, so that their evolution is mostly determined by internal processes rather than their environment.

In summary, after controlling for mass, we find strong evidence for environmental influence in the chemical and star-formation history of disk galaxies, particularly in low-mass systems.  We can also quantify these differences in terms of the underlying physical processes of pristine gas accretion and processed gas loss, in the sense that low-mass satellites have their gas accretion shut off at an earlier stage, and are more likely to lose their processed gas rather than recycling it into further generations of stars.  We also find a clear difference between central and isolated galaxies, in that central galaxies continue to accrete gas over longer timescales, and are better at holding on to that gas once processed through a generation of stars.  We note that these physically-plausible differences were not at all imposed by the semi-analytic spectral fitting process, which had no way of knowing whether a particular spectrum came from a central, satellite, or isolated galaxy.  The differences therefore seem to be astrophysical in origin, and offer further confirmation of the reliability of this archaeological approach to studying galaxy evolution.

\section{Conclusions}
In this work we have applied the semi-analytic spectral fitting approach to a sample of disk galaxies selected from the SDSS-IV/MaNGA survey to investigate the dependence of such galaxies' evolution on their environment. The sample comprised three mass-matched sets of isolated, central, and satellite galaxies, the latter two categories defined by whether or not the galaxy is the most massive member of a group.  The data for each galaxy within one effective radius were combined to create a high signal-to-noise ratio spectrum representing the system's global properties.  The semi-analytic spectral fitting process not only reproduces the properties of the spectrum by fitting a star-formation and chemical evolution history, but also determines physical parameters of gas accretion and outflow that generate the model.  From these fits, we have determined that:

\begin{itemize}

\item
The cumulative SFHs derived from the sample galaxies reveal
an earlier cessation of star formation activities in low-mass ($10^{9.5}<M_*/{\rm M}_{\odot}<10^{10.0}$) 
satellite galaxies than their central equivalents: less than 40\% of the stellar masses in satellite galaxies formed within the most recent 5\,Gyrs, while central galaxies of similar mass accumulated more than 70\% of their stellar masses over the same period.  Isolated galaxies have intermediate properties.  This environmental dependence becomes weaker in more massive galaxies, and essentially disappears for systems with $M_*/{\rm M}_{\odot}>10^{10.5}$. These differences also fit with a more qualitative analysis looking at the abundances of $\alpha$ elements in different galaxies, with low-mass satellites showing the sort of $\alpha$ enhancement associated with a briefer period of star formation.

\item
The environmental dependence of disk galaxy evolution, as well as its mass dependence, is also seen in the chemical evolution of the galaxies. Low-mass ($10^{9.5}<M_*/{\rm M}_{\odot}<10^{10.0}$) satellite galaxies contain a high fraction of low metallicity stars compared to central galaxies, indicating that they are more likely to be `closed-box' or `leaky-box' systems, while central galaxies of similar masses contain lower fractions of low metallicity stars, suggesting that on-going accretion played a more important role in their formation.  Once again, isolated galaxies lie between central and satellite galaxies, and environmental dependence becomes negligible in galaxies more massive than $\sim10^{10.5}{\rm M}_{\odot}$.

\item
The gas accretion history inferred from the semi-analytic fitting process drives the variations in timescale: in low-mass central galaxies, the average gas infall timescale is found to be systematically longer than comparable satellite galaxies, leading to their more extended star formation histories.  Again, this variation disappears in more massive galaxies, where we also start to see a different sub-population of S0 galaxies, whose gas was accreted over a very short timescale.  However, the timescales are not that short for the low-mass satellite systems, suggesting the slow `strangulation' of their star formation caused by the removal of their extended hot halo of gas rather than more abrupt ram-pressure stripping.

\item
This scenario is further supported by the inferred mass-loss parameters in low-mass galaxies, where satellite systems are found to have stronger wind parameters, indicating that their processed gas has been effectively removed.  By contrast, central galaxies have weaker wind parameters than their isolated equivalents, as would be expected if the deeper potential well of a group were better at retaining such processed material.

\end{itemize}

The consistent story obtained from the semi-analytic spectral fitting process offers real confidence that even the simple chemical evolution model adopted, with its parameterisation of gas accretion and mass loss, provides a credible physical description of the histories of these galaxies.  This new direct model fitting approach, combined with the large sample size and high quality data in the MaNGA survey, have revealed the significant impact of environment on lower mass galaxies, and allowed us to quantify it in a way that has not been possible previously.

\section*{Acknowledgements}
SZ, MRM and AAS acknowledge financial support from the UK Science and Technology Facilities Council (STFC; grant ref: ST/T000171/1).

For the purpose of open access, the authors have applied a creative commons attribution (CC BY) to any journal-accepted manuscript.

Funding for the Sloan Digital Sky Survey IV has been provided by the Alfred P. 
Sloan Foundation, the U.S. Department of Energy Office of Science, and the Participating Institutions. 
SDSS-IV acknowledges support and resources from the Center for High-Performance Computing at 
the University of Utah. The SDSS web site is www.sdss.org.

SDSS-IV is managed by the Astrophysical Research Consortium for the Participating Institutions of the SDSS Collaboration including the Brazilian Participation Group, the Carnegie Institution for Science, Carnegie Mellon University, the Chilean Participation Group, the French Participation Group, Harvard-Smithsonian Center for Astrophysics, Instituto de Astrof\'isica de Canarias, The Johns Hopkins University, Kavli Institute for the Physics and Mathematics of the Universe (IPMU) / University of Tokyo, Lawrence Berkeley National Laboratory, Leibniz Institut f\"ur Astrophysik Potsdam (AIP), Max-Planck-Institut f\"ur Astronomie (MPIA Heidelberg), Max-Planck-Institut f\"ur Astrophysik (MPA Garching), Max-Planck-Institut f\"ur Extraterrestrische Physik (MPE), National Astronomical Observatories of China, New Mexico State University, New York University, University of Notre Dame, Observat\'ario Nacional / MCTI, The Ohio State University, Pennsylvania State University, Shanghai Astronomical Observatory, United Kingdom Participation Group, Universidad Nacional Aut\'onoma de M\'exico, University of Arizona, University of Colorado Boulder, University of Oxford, University of Portsmouth, University of Utah, University of Virginia, University of Washington, University of Wisconsin, Vanderbilt University, and Yale University.

\section*{Data availability}
The data underlying this article were accessed from: SDSS DR17 \url{https://www.sdss.org/dr17/manga/}. The derived data generated in this research will be shared on request to the corresponding author.

\label{sec:summary}
\bibliographystyle{mnras}
\bibliography{szhou} 

\bsp	
\label{lastpage}
\end{document}